\documentclass{article}
\usepackage{amsmath,amssymb, amsthm, amsbsy,bm}
\usepackage{subcaption}
\usepackage{graphicx}
\usepackage{booktabs}
\usepackage{xcolor}
\begin{document}

\title{Phase field modelling of the growth and detachment of bubbles in a hydrogen electrolizer}
\author{Carlos Uriarte \\
\'{A}rea de Electromagnetismo, Universidad Rey Juan Carlos, \\
Tulip\'{a}n s/n, Mostoles, 28933, Madrid, Spain \and Marco A. Fontelos \\
Instituto de Ciencias Matem\'{a}ticas (ICMAT, CSIC-UAM-UC3M-\\
UCM), C/ Nicol\'{a}s Cabrera 15, 28049 Madrid, Spain. \and Manuel Array\'{a}%
s \\
\'{A}rea de Electromagnetismo, Universidad Rey Juan Carlos, \\
Tulip\'{a}n s/n, M\'{o}stoles, 28933, Madrid, Spain}
\maketitle

\begin{abstract}
We build and solve numerically a phase field model for the growth and detachment of a gas bubble resting on an electrode and being filled with hydrogen produced by water electrolysis. The bubble is surrounded by a viscous liquid, has a prescribed static contact angle and is also subject to gravitational forces. We compute, as a function of the static contact angle, the time at which the bubble detaches from the substrate and what volume it has at that time. We also investigate the dependence of the detachment time on other parameters such as the applied voltage and the hydrogen ion concentration at the fluid bulk.
\end{abstract}

\section{Introduction}
The study of the nucleation, growth and detachment of gas bubbles inside viscous liquids has relevance across a wide range of applications including boiling (cf.~\cite{Z}), cavitation (cf.~\cite{FA}), microfluidics  (cf.~\cite{JO}), and electrochemical systems (cf.~\cite{ZLF}). In the electrochemical context, gas bubbles form at electrodes as a result of chemical reactions, such as the hydrogen generation during water electrolysis, a process of growing industrial importance for sustainable energy storage and conversion (see for instance \cite{YL}). Hence, the search for methods and techniques to optimize the energy production (cf.~\cite{YY}) becomes of special interest.

In our previous work \cite{US}, we have investigated the dynamics of bubble detachment under different external conditions using a phase field approach (for a general description of the method see \cite{C}, \cite{DF},  for its application for fluid mechanical problems \cite{J}, \cite{VCB}, \cite{TQ}, \cite{EFG}, \cite{FGK} and for the implementation in a problem for bubble detachment \cite{AFU}). This method, which replaces sharp interfaces with smoothly varying diffuse interfaces, allowed us to effectively model the complex interfacial dynamics and topological changes associated with bubble evolution. We focused on the interplay between buoyancy, surface tension, and hydrodynamic forces in determining the detachment behaviour of bubbles, providing insights into the physical mechanisms governing gas release in electrochemical systems.

\begin{figure}[t]
\includegraphics[width=1.0\textwidth]{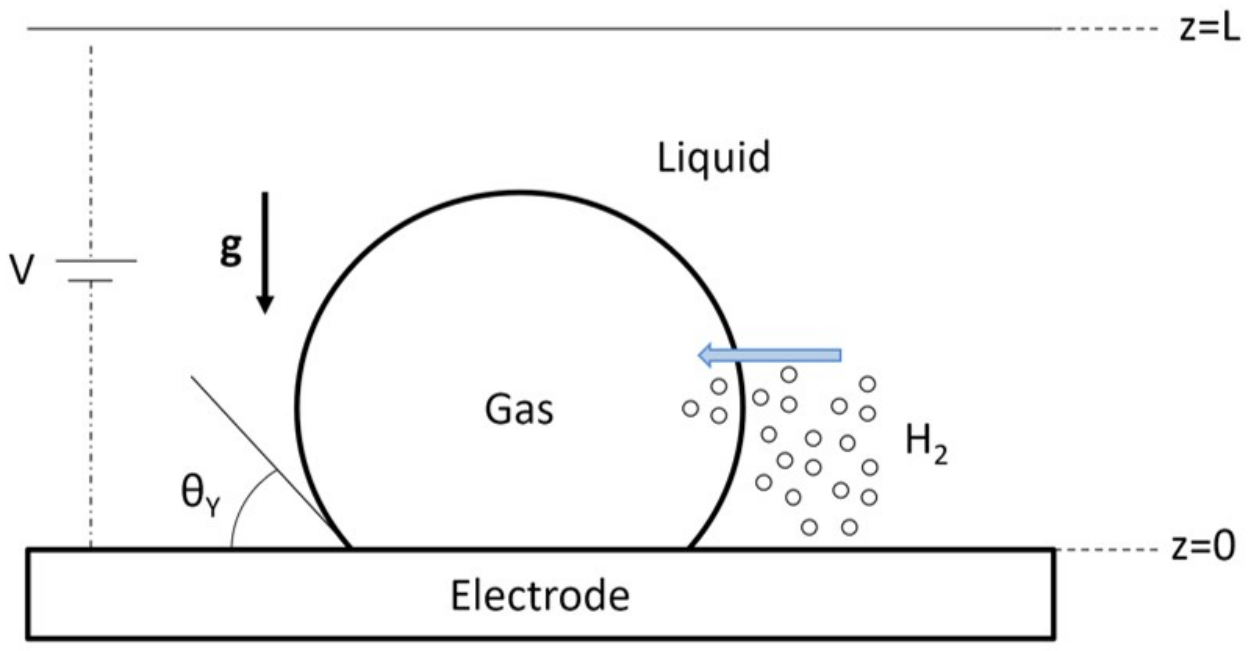}
\caption{Sketch of physical settings: H$_2$ is produced at the electrode by water electrolysis using the electrons $e^-$  provided by the battery, and diffuses inside the bubble with the contact angle $\theta_Y$. The bubble will grow and detach once buoyancy forces due to gravity overcome the surface tension forces that attach the bubble to the surface.}
\label{fig1}
\end{figure}

In the present study, we extend this framework by incorporating the effects of electrochemical reaction kinetics and electric fields. These additions are essential for a more comprehensive understanding of bubble dynamics in realistic electrolysis environments, where the local gas production rate and the electric field distribution significantly influence bubble growth and detachment. By coupling the phase field model with reaction-diffusion equations and electrostatic interactions, we aim to capture the feedback between electrochemical activity and fluid dynamics, offering a more complete picture of gas evolution at electrodes.

We remark that, in contrast to \cite{US}, the situation described here is much more complex to model. The presence of electrodes where reaction takes place and of chemical species with nontrivial boundary conditions at the surface of the bubble, introduce extra difficulties. We will implement the boundary conditions by suitably coupling the phase field with the species transport equations.  The role of chemical reactions at the electrode is not only to provide the gas that fills the bubble; but to change the global geometry of the problem. As the bubble expands, including the gas/solid contact región, the área of the electrode where reaction takes place is modified and hence the production of hydrogen is influenced by the dynamics of the bubble itself. This makes the results to deviate from our previous paper due to this non trivial coupling between chemical reaction and fluid dynamics.

The bubble grows due to the influx of hydrogen molecules (H$_{2}$) that is produced through a water electrolysis process at the cathode:%
\[
2\text{H}_{2}\text{O}\ +\ 2\ \text{e}^{-}\rightarrow \text{H}_{2}+2\ \text{OH}^{-}.
\]
In figure \ref{fig1}, we provided a sketch of the physical situation to be considered in this work. The phase field model coupled with Navier-Stokes equation which we used in our previous work \cite{US} must be extended in order to incorporate the electrochemistry processes. Sections 2, 3, and 4 are dedicated to the development of this extended model. In Section 2, we introduce the modelling of electrochemical phenomena and the motion of the involved fluids. In section 3 some dimensionless parameters are introduced, along with their corresponding physical values. Section 4 focuses on incorporating the phase-field formulation into the previously established equations in order to develop a complete diffuse interface model to be implemented numerically.  Section 5 contains the numerical simulations of the model. In particular, we investigate the influence of the contact angle on the detachment time and gas bubble volume, as well as the effect of the applied electrode potential on the overall process. The paper concludes with a summary of our findings and prospects for future work.

\section{Mathematical model}

In this section we will develop a mathematical model for the production of
hydrogen at an electrolytic device and the growth of a bubble filled by
gaseous hydrogen. The model will entail essentially two parts: 1) the motion
of the fluids involved, which will typically be water and gaseous hydrogen,
2) the production and flux of hydrogen molecules from the reaction at the
electrode, as well as hydrogen ions (protons).

\subsection{Fluid equations}

Both the external flow and the growing bubble are fluids. In the first case
we typically consider water and, in the second, gaseous hydrogen. The
velocities involved are small in comparison with the sound velocity so that
we will neglect compressibility effects of the gas inside the bubble and
will merely treat it as an incompressible fluid. Hence, denoting $\rho _{i}$%
, $i=1,2$ the densities of both fluids, the velocity field will satisfy
Navier-Stokes equations:
\begin{eqnarray}
\frac{\partial (\rho _{i}\mathbf{v)}}{\partial t}+\rho _{i}\mathbf{v}\cdot
\nabla \mathbf{~v}-\nabla \cdot \mathbf{S}_{i} &=&-\nabla p-\rho _{i}g%
\mathbf{e}_{z}, \\
\nabla \cdot \mathbf{v} &=&0,
\end{eqnarray}
with a viscous stress
\begin{equation}
\mathbf{S}_{i}=\frac{\eta _{i}}{2}\left( \nabla \mathbf{v}+\nabla \mathbf{v}%
^{T}\right) 
\end{equation}%
where $\eta_{i}$, $i=1,2$ are the viscosities of both fluids, and the
balance of force condition at the interface:
\begin{equation}
\left( -p\mathbf{I}+\mathbf{S}_{i}\right) \cdot \mathbf{n},
=\sigma \kappa \mathbf{n} 
\end{equation}%
being $\kappa (\mathbf{x},t)$ the mean curvature at any point $\mathbf{x}$
at the interface between the bubble and the surrounding liquid. We take as
initial configuration for the bubble a truncated sphere with radius $R$.

We introduce a characteristic velocity $U$ defined by balancing inertial and gravitational forces:
\[
\rho _{1}\frac{U^{2}}{R}=\rho _{1}g,
\]%
which allows us to write the system in terms of two dimensionless numbers:
\begin{equation}
Bo=\frac{gR^{2}\rho _{1}}{\sigma },\,\,\, Re=\frac{UR\rho_1}{\eta_1} ,
\end{equation}
where $Bo$ is the Bond number and ${Re}$ the Reynolds number in the liquid phase ($\rho_1$). Finally, defining the density and viscosity ratios%
\begin{equation}
\gamma =\frac{\rho _{2}}{\rho _{1}},\ \chi =\frac{\eta_{2}}{\eta _{1}},
\end{equation}%
we conlcude the following equations for the pressure and the velocity field:%
\begin{eqnarray}
\frac{\partial \mathbf{v}}{\partial t}+\mathbf{v}\cdot \nabla \mathbf{~v}-%
\frac{1}{Re}\Delta \mathbf{v} &=&-\nabla p-Bo\ \mathbf{e}_{z},\ \ \text{%
in the liquid phase} ,\\
\frac{\partial \mathbf{v}}{\partial t}+\mathbf{v}\cdot \nabla \mathbf{~v}-%
\frac{\chi }{Re}\Delta \mathbf{v} &=&-\nabla p-\gamma Bo\ \mathbf{e}%
_{z},\ \text{in the gas phase}.
\end{eqnarray}

\subsection{Electrochemistry equations and 1D stationary
solutions}
The concentration $n$ of protons ($\text{H}^{+}$), $n_{\text{H}_{2}}$ of $\text{H}_{2}$ and the electric potential $\varphi $ satisfy under stationary conditions in the bulk%
\begin{eqnarray}
\Delta n &=&0, \label{eq:1}\\
\Delta n_{\text{H}_{2}} &=&0, \\
\nabla \cdot (n\nabla \varphi ) &=&0 \label{eq:3}.
\end{eqnarray}%
Far from the cathode, at $z=L$, we can take some reference values,
\begin{equation}
  n=n_{r},\ n_{\text{H}_{2}}=n_{\text{H}_{2r}},\ \varphi =V, \text{ at }z=L.
  \label{eq:xL}
\end{equation}
At the cathode ($z=0$) we have the following conditions in the region of the electrode surface not covered by bubbles,
\begin{eqnarray}
\frac{\partial n}{\partial z} &=&\frac{1}{D_{+}}\omega, \label{1}\\
\frac{\partial n_{\text{H}_{2}}}{\partial z} &=&-\frac{1}{D_{\text{H}_{2}}}\omega, \label{2}\\
\frac{F}{RT}\frac{\partial \varphi }{\partial z} &=&\frac{1}{n}\omega,\label{3}
\end{eqnarray}%
where the reaction rate $\omega $ satisfies the Butler-Volmer equation
\begin{equation}
\omega =\frac{i_{0}}{2F}\left( e^{\frac{2\alpha F}{RT}(\varphi -\varphi
_{r})}\left( \frac{n}{n^{s}}\right) ^{2}-e^{-\frac{2(1-\alpha )F}{RT}%
(\varphi -\varphi _{r})}\frac{n_{\text{H}_{2}}}{n^{s}}\right) ,
\end{equation}%
with $n^{s}$ is some reference concentration (say $1$ mol/l for instance)
and $\varphi _{r}$ is some reference potential. Here $\omega$ is the amount of protons produced per unit time and unit area. $R$, $F$ and $T$ are the universal gas constant, the Faraday constant and the temperature respectively. The transfer coefficient $\alpha$ is a model constant, while $i_{0}$  is the exchange current in the standard reference state, which is a function of the material and structure of the electrode. The conditions \eqref{1}-\eqref{3} express the balance of the flux of each electro active species entering or leaving the double layer -- formed at the cathode -- by diffusion and migration from the bulk. In equations \eqref{1} and \eqref{2}, $D_{+}$ and $D_{\text{H}_{2}}$ are  the diffusivity of protons and dissolved hydrogen respectively. The electric field at the boundary layer is given by \eqref{3} (further details of the model can be found in \cite{Modelo}). 

We have the following 1D solution for the system \eqref{eq:1}-\eqref{eq:3},
\begin{eqnarray*}
\frac{\partial \varphi}{\partial z} &=&\frac{K}{n(z)}, \\
n(z) &=&A+Bz, \\
n_{\text{H}_{2}}(z) &=&C+Dz,
\end{eqnarray*}%
and hence%
\begin{eqnarray}
\varphi (z) &=&\frac{K}{B}\log (A+Bz)+E,
\end{eqnarray}
where $K, A, B, C, D$ and $E$ are constants to be determined from boundary conditions. 
The conditions \eqref{eq:xL} at $z=L$ yield,
\begin{eqnarray}
V &=&\frac{K}{B}\log (A+BL)+E, \\
n_{\text{H}_{2r}} &=&C+DL, \\
n_{r} &=&A+BL.
\end{eqnarray}%
On the other hand, from the boundary conditions at the electrode we have%
\begin{eqnarray}
-D_{\text{H}_{2}}D &=&\omega,  \\
B &=&\frac{1}{D_{+}}\omega,  \\
\frac{F}{RT}\frac{K}{A} &=&\frac{1}{AD_{+}}\omega. 
\end{eqnarray}%
Solving the system we get%
\begin{eqnarray}
D &=&-\frac{1}{D_{\text{H}_{2}}}\omega,  \\
C &=&n_{\text{H}_{2r}}+\frac{L}{D_{\text{H}_{2}}}\omega,  \\
B &=&\frac{1}{D_{+}}\omega,  \\
A &=&n_{r}-\frac{L}{D_{+}}\omega,  \\
K &=&\left( \frac{F}{RT}\right) ^{-1}\frac{1}{D_{+}}\omega,  \\
E &=&\left( \frac{F}{RT}\right) ^{-1}\log n_{r}-V,
\end{eqnarray}%
all in terms of the reaction rate at the electrode $\omega$. We can then
introduce in the Butler-Volmer equation the concentrations and potential as
functions of $\omega $ and obtain an algebraic equation for $\omega $. In
fact, we can find an equilibrium (no production of hydrogen) when%
\begin{equation}
e^{-2\frac{F}{RT}\left( \varphi _{r}-\varphi _{eq}\right) }=\frac{n_{r}^{2}}{
n_{\text{H}_{2r}}}.
\end{equation}%
Then one has, as a function of the value of the potential at the electrode $%
\varphi $ 
\begin{equation}
\omega =\frac{i_{0}}{2F}\left[ \left( \frac{n_{r}^{2}}{n_{\text{H}_{2r}}}\right)
^{-\alpha }e^{\frac{2\alpha F}{RT}(\varphi -\varphi _{r})}n^{2}-e^{-\frac{%
2(1-\alpha )F}{RT}(\varphi -\varphi _{r})}\left( \frac{n_{r}^{2}}{n_{\text{H}_{2r}}}%
\right) ^{1-\alpha }n_{\text{H}_{2}}\right] ,
\end{equation}
and since%
\begin{eqnarray*}
\varphi (0)-\varphi _{r} &=&\frac{K}{B}\log  A+E=V+\left( \frac{F}{RT}%
\right) ^{-1}\log (1-\frac{L}{n_{r}D_{+}}\omega ) ,\\
n(0) &=&A=n_{r}-\frac{L}{D_{+}}\omega , \\
n_{\text{H}_{2}}(0) &=&C=n_{\text{H}_{2r}}+\frac{L}{D_{\text{H}_{2}}}\omega ,
\end{eqnarray*}%
we obtain the following relations for small deviations from the equilibrium $\delta \varphi, \delta\omega, \delta V$:%
\begin{eqnarray}
\delta \varphi  &=&\delta V-\left( \frac{F}{RT}\right) ^{-1}\frac{L}{%
n_{r}D_{+}}\delta \omega,  \\
\delta \omega  &=&\frac{i_{0}}{2F}\left[ \frac{2F}{RT}\left( \frac{n_{r}^{2}%
}{n_{\text{H}_{2r}}}\right) ^{1-\alpha }n_{\text{H}_{2r}}\right] \delta \varphi \nonumber \\
&&+\frac{i_{0}}{2F}\left[ \left( \frac{n_{r}^{2}}{n_{\text{H}_{2r}}}\right)
^{-\alpha }2n_{r}\left( -\frac{L}{D_{+}}\delta \omega \right) -\left( \frac{%
n_{r}^{2}}{n_{\text{H}_{2r}}}\right) ^{1-\alpha }\frac{L}{D_{\text{H}_{2}}}\delta \omega
\right], 
\end{eqnarray}%
leading to%
\begin{equation}
\delta V=\frac{1+\frac{i_{0}}{F}\frac{2L}{D_{+}}\left( \frac{n_{r}^{2}}{%
n_{\text{H}_{2r}}}\right) ^{1-\alpha }\frac{n_{\text{H}_{2r}}}{n_{r}}+\frac{i_{0}}{2F}%
\left( \frac{n_{r}^{2}}{n_{\text{H}_{2r}}}\right) ^{1-\alpha }\frac{L}{D_{\text{H}_{2}}}}{%
\frac{i_{0}}{2F}\left( \frac{2F}{RT}\left( \frac{n_{r}^{2}}{n_{\text{H}_{2r}}}%
  \right) ^{1-\alpha }n_{\text{H}_{2r}}\right) }\delta \omega,
\label{eqvar}
\end{equation}
which yields a linear relation between the reaction rate $\delta \omega$ and the variations in the applied over potential $\delta V$. Note that prefactor of $\delta \omega$ in \eqref{eqvar} is positive, so that an increment in the potential corresponds to an increment on the reaction rate. 

\subsection{Bubble H$_{2}$ influx}

The influx of $\text{H}_{2}$ that fills the bubble is given by%
\begin{equation}
  J=-\int_{\partial \Omega }D_{\text{H}_{2}}\nabla n_{\text{H}_{2}}\cdot \mathbf{n}\,dS,
  \label{eq:influx}
\end{equation}
where $\partial \Omega $ is the part of the surface of the bubble that is in
contact with the surrounding liquid. If we assume that the temperature $T$
remains constant, the concentration of $\text{H}_{2}$ is constant throughout the
bubble and the gas satisfies the law of perfect gases. We have then for a given mass $m$ of $\text{H}_{2}$
\begin{equation}
pV=n_{\text{H}_{2}}RT=\frac{m}{m_{\text{H}_{2}}}RT,
\end{equation}
being $m_{\text{H}_{2}}$ the H$_2$ molecular mass, which allows to compute the volume of the bubble at any given time knowing
the pressure and the mass $m$. Taking time derivatives we deduce 
\[
\frac{dp}{dt}V+p\frac{dV}{dt}=\frac{RT}{m_{\text{H}_{2}}}\frac{dm}{dt},
\]%
so the change in the volume is 
\begin{equation}
  \frac{dV}{dt}=\frac{RT}{m_{\text{H}_{2}}p}\left(\frac{dm}{dt}-m\frac{d\log p}{dt}\right).
\end{equation}
Using \eqref{eq:influx}, the total mass can be computed at any time since
\begin{equation}
  \frac{dm}{dt}=m_{\text{H}_{2}}J.
 \end{equation} 
The pressure $p$ can be found by solving the hydrodynamic equations.
Nevertheless, we can safely assume the term involving $dp/dt$ to be
small compared with other terms since the pressure inside the bubble, for a bubble which grows slowly, is determined from the balance between jump in pressure across the interface
and surface tension (i.e. the inverse of the bubble radius) and this changes slowly in time.  Hence, the change in volume will be determined from the change in the total mass of $\text{H}_{2}$ inside the bubble. More precisely, we assume that the mass of $\text{H}_{2}$ produced at the electrode diffuses inside the bubble. Thus, the amount of gas across the boundaries is neglected since it is very small compared to the amount of gas produced. We will verify this by direct numerical simulations. We will inject the corresponding amount of mass per unit time as a source inside the bubble.

\section{Physical parameters and scales}

In this section we discuss the typical values of the physical parameters
involved in the problem. Firstly, concerning the fluid parameters (for water and H$_2$) we
have at 298 K and 1 atm,%
\begin{eqnarray}
\rho _{1} &=&10^{3}\ \text{kg/m}^{3},\nonumber \\
\rho _{2} &=&0.08988\ \text{kg/m}^{3}, \nonumber\\
\eta _{1} &=&0.89\cdot 10^{-3}\ \text{Pa}\cdot \text{s},\label{parameters} \\
\eta _{2} &=&8.874\cdot 10^{-6}\ \text{Pa}\cdot \text{s} ,\nonumber\\
\sigma  &=&72.8\cdot 10^{-3}\ \text{N/m}\nonumber.
\end{eqnarray}
Secondly, we note that, at 298 K and 1 atm,%
\[
\frac{F}{RT}=\frac{96500\ \text{C/mol}}{(8.314\ \text{J/(mol  K)})\cdot (298\ \text{K})}%
\allowbreak =38.\,\allowbreak 95\ \text{V}^{-1}.
\]%
We can also introduce the dimensionless number
\begin{equation}
\iota =\frac{i_{0}L}{FD_{H_{2}}n_{s}}=0.83,
\end{equation}
by taking for instance $L=0.01\ \text{m}$ and $i_{0}=10^{-2}$ A/m$^{2}$, and using \eqref{parameters}.
In addition, the diffusivity of hydrogen and $\text{H}^{+}$ in water are%
\begin{eqnarray}
D_{\text{H}_{2}} &=&1.6\times 10^{-9}\ \text{m}^2/\text{s},\\
D_{+} &=&9.3\times 10^{-9}\ \text{m}^2/\text{s},
\end{eqnarray}
respectively,  and we can define the diffusivity ratio as
\begin{equation}
\delta =\frac{D_{\text{H}_{2}}}{D_{+}}=\frac{1.6}{9.31}=0.17.
\end{equation}
Next we rescale the concentrations with the saturation concentration for $\text{H}_{2}$ 
\begin{equation}
n_{s}=7.8\times 10^{-4}\ \text {mol/l},
\end{equation}
so that we make concentrations dimensionless and rescale the potential with $%
(\frac{F}{RT})^{-1}$ to make it dimensionless. We rescale space with the length 
$L$ and finally get the rescaled boundary conditions,
\begin{eqnarray}
\frac{\partial \varphi }{\partial z} &=&\frac{1}{n}\delta \omega  \\
\frac{\partial n}{\partial z} &=&\delta \omega  \\
  \frac{\partial n_{\text{H}_{2}}}{\partial z} &=&-\omega 
\end{eqnarray}%
at the electrode. Similarly%
\begin{equation}
n=n_{r},\ n_{\text{H}_{2}}=n_{\text{H}_{2r}},\ \varphi =V\ \ \ \text{at }z=1
\end{equation}%
The Butler-Volmer equation is then
\begin{equation}
\omega =\frac{\iota }{2}\left( e^{2\alpha (\varphi -\varphi
_{eq})}n^{2}-e^{-2(1-\alpha )(\varphi -\varphi _{eq})}n_{\text{H}_{2}}\right) .
\end{equation}
Note that $\omega =0$ (no reaction) when $\varphi =\varphi _{0}$ with%
\[
\varphi _{0}-\varphi _{eq}=\frac{1}{2}\log \frac{n^{2}(z=0)}{n_{\text{H}_{2}}(z=0)} ,
\]%
and since $n=n_{r}$, $n_{\text{H}_{2}}=n_{\text{H}_{2r}}$, $V=\varphi _{0}$ in that case
we have%
\[
\varphi _{0}-\varphi _{eq}=\frac{1}{2}\log \frac{n_{r}^{2}}{n_{\text{H}_{2r}}}.
\]

If we take $V>\varphi _{0}$ then $n_{\text{H}_{2}}(z)\geq n_{\text{H}_{2r}}$, $n(z)\leq
n_{r}$. We will typically take $n_{\text{H}_{2r}}=1$ close to saturation so that
the bulk is over saturated, $n>1$ (so that the concentration of $\text{H}^{+}$ is
larger than the saturation concentration which is under standard conditions $%
7.8\cdot 10^{-4}$ mol/l that would represent a $\text{pH}$ of $3.1$).

\section{Phase field modelling}

One of the most powerful approaches for studying multiphase flows is the phase field method. This method replaces sharp interfaces between fluids with diffuse interfaces, where a phase field function, denoted by $\phi (\mathbf{x},t)$, undergoes rapid transitions across a thin interfacial region of thickness approximately $\varepsilon$ (with $\varepsilon$ being sufficiently small). The function $\phi$ typically varies between two limiting values (e.g., $\phi = 1$ and $\phi = -1$), each representing a different fluid phase.

A key advantage of using diffuse interfaces is their natural ability to accommodate topological changes in the fluid domain, such as merging or breaking of fluid regions. To model this behavior, one must define an appropriate partial differential equation (PDE) for the phase field function and couple it consistently with the other fluid variables. This topic has been extensively studied in the literature (see, for example, \cite{J,VCB,TQ,EFG,FGK}).

The most suitable governing equation for the phase field function is a fourth-order PDE known as the Cahn–Hilliard equation, which was first introduced in \cite{CH} (see also \cite{AC}). In the context of multiphase flows, this equation is modified to include a convection term, involving the fluid velocity $\mathbf{v}(\mathbf{x},t)$, to accurately capture the transport of the phase field:

\begin{equation}
\frac{\partial \phi }{\partial t}+\mathbf{v}\cdot \nabla \phi =\nabla \cdot
\left( M\nabla \mu \right),  \label{b1}
\end{equation}%
with%
\begin{equation}
\mu =-\varepsilon \Delta \phi +\frac{1}{\varepsilon }W^{\prime }(\phi ),
\label{b2}
\end{equation}%
where $\mu $ is the chemical potential, $M$ is a ``mobility" factor $W(\phi )$ is a
phase-field potential having two local minima at the values
of $\phi $ that correspond to the two phases. We will take specifically,
\begin{equation}
W(\phi )=\phi ^{2}(1-\phi ^{2}).
\end{equation}
\subsection{Navier-Stokes}

We introduce the phase field in the Navier-Stokes system as
\begin{equation}
\frac{\partial (\rho \left( \phi \right) \mathbf{v)}}{\partial t}+\rho
\left( \phi \right) \mathbf{v}\cdot \nabla \mathbf{~v}-\nabla \cdot \mathbf{S%
}=-\nabla p+\mu \nabla \phi -\rho \left( \phi \right) g\mathbf{e}_{z}, 
\end{equation}
where the viscous stress tensor is%
\begin{equation}
\mathbf{S}=\frac{\eta \left( \phi \right) }{2}\left( \nabla \mathbf{v}+\nabla 
\mathbf{v}^{T}\right).
\end{equation}
The material parameters $\rho \left( \phi \right) $ and $\eta \left( \phi
\right) $ are linear interpolation between the fluids densities and viscosities:

\begin{eqnarray}
\rho \left( \phi \right) &=&\rho _{1}\frac{1-\phi}{2} +\rho _{2}\frac{1+\phi}{2}, \\
\eta \left( \phi \right) &=&\eta _{1}\frac{1-\phi}{2} +\eta_{2}\frac{1+\phi}{2}.
\end{eqnarray}%
Also fluid incompressibility will be imposed,%
\begin{equation}
\nabla \cdot \mathbf{v}=0.
\end{equation}
Concerning the phase field boundary conditions, we first take
\begin{equation}
\frac{\partial \mu }{\partial n}=0  \label{b3},
\end{equation}%
so that there is no flux of chemical potential through the boundary of the domain. Secondly, following \cite{EFG}
and \cite{TQ}, we additionally impose the condition 
\begin{equation}
\sigma _{0\text{ }}\varepsilon \frac{\partial \phi }{\partial n}=\sigma
_{fs}^{\prime }(\phi )  \label{b4},
\end{equation}%
where $\sigma _{fs}^{\prime }(\phi )$ interpolates between the liquid/solid
interfacial energy $\sigma _{LS}$ and the gas/solid interfacial energy $%
\sigma _{GS}$ by means of the following formula:%
\begin{equation}
\sigma _{fs}(\phi )=\frac{\sigma _{GS}+\sigma _{LS}}{2}+\frac{\sigma
_{GS}-\sigma _{LS}}{2}\sin \left(\frac{\pi \phi }{2}\right).
\end{equation}
$\sigma _{0}$ depends on the liquid/gas interfacial energy $%
\sigma _{LG}$ as:%
\begin{equation}
\sigma _{0}=\frac{3\sqrt{2}}{8}\sigma _{LG},
\end{equation}
so that the condition (\ref{b4}) imposes, in the limit $\varepsilon
\rightarrow 0$, a liquid/gas contact (or Young's) angle $\theta _{Y}$ given by
\begin{equation}
\cos \theta _{Y}=\frac{\sigma _{GS}-\sigma _{LS}}{\sigma _{GS}} .
\end{equation}

We remark that the condition \eqref{b4} also incorporates, in the case when the contact line is moving, and in the limit $\varepsilon
\rightarrow 0$, classical models for contact line motion such as Cox-Voinov as demostrated in several studies such as \cite{FENG}.

\subsection{Electrochemistry}

We introduce the phase field into the equations for the concentration of $%
\text{H}_{2}$ by writing the diffusion coefficient as a function of the phase field $\phi $:%
\begin{equation}
D_{1}(\phi )=D_{\text{H}_{2}}\frac{1+\phi }{2} .
\end{equation}
We impose $n_{\text{H}_{2}}=n_{s}$ at the surface of the bubble by introducing a
penalization term $C_{b}\left( n_{s}-n_{\text{H}_{2}}\right) $ with $C_{b}\gg 1$ so
that $n_{\text{H}_{2}}\simeq n_{s}$ inside the bubble and hence at its boundary.
The resulting equation is
\begin{equation}
  \nabla \cdot \left( D_{1}(\phi )\nabla n_{\text{H}_{2}}\right) +C_{b}\left(
n_{s}-n_{\text{H}_{2}}\right) =0. 
\end{equation}
Analogously, for the concentration of $\text{H}^{+}$ and the electrostatic
potential we write the equations%
\begin{equation}
D_{2}(\phi )=D_{{+}}\frac{1+\phi }{2},\ \ \ \nabla \cdot (D_{2}(\phi )\nabla n)=0,\ \ \ 
\end{equation}
and
\begin{equation}
\nabla \cdot (n\nabla \varphi )=0,
\end{equation}
respectively. Finally, the reaction rate at the electrode is corrected in
such a way that no reaction takes place at the part of the electrode covered
by the bubble:
\begin{equation}
\omega _{\phi }=\frac{1+\phi }{2}\omega.
\end{equation}
Thus the conditions at the reacting electrode are%
\begin{eqnarray}
\frac{\partial n}{\partial z} &=&\frac{1}{D_{+}}\omega _{\phi }, \\
\frac{\partial n_{\text{H}_{2}}}{\partial z} &=&-\frac{1}{D_{\text{H}_{2}}}\omega _{\phi },
\\
\frac{F}{RT}\frac{\partial \varphi }{\partial z} &=&\frac{1}{n}\omega _{\phi},
\end{eqnarray}%
and the conditions at $z=L$ remain unchanged:%
\begin{equation}
n=n_{r},\ n_{\text{H}_{2}}=n_{\text{H}_{2r}},\ \varphi =V.
\end{equation}

\section{Numerical results}

We will use the phase field model developed in the previous sections to simulate the behaviour of a growing bubble under various physical conditions. First we will study the effect of the contact angle on time of detachment and the volume of gas detached with the bubble. Secondly, we will see how the magnitude of the applied potential $V$ influences the gas production.

The simulations have been carried out using the Comsol PDE software \cite{COM}. The mathematical models previously proposed have been introduced and solved using a finite element scheme, coupling the different physics through a semi-discrete time approach. This approach divides the governing equations into the time-independent (diffusion and potential) and the time-dependent equations (phase field model and Navier Stokes, NS from now on, equations) in such a way that the time domain is subdivided into $n$ distinct intervals of $0.1$ s duration each, between which the diffusion physics of $n_{\text{H}_{2}}$, $n_{p}$ and the potential are recalculated. Therefore, at each time period of $0.1$ s, the phase field and velocity field of the NS equations are solved, using stationary values (calculated previously) of the concentrations and potential during this time. After this time, the values of the concentrations are recalculated taking into account the values of the phase field obtained in the last instant of time calculated in the previous step.

In our simulations we use a cylindrical system of coordinates and we will assume axial symmetry with the origin placed in the electrode plane, being $z$ and $r$ the vertical and radial coordinates respectively. The simulation domain consists of a cylinder supported at z=0, with 3 cm radius and 1 cm height (simulating an electrolysis cell with a circular electrode of 6 cm diameter and 1 cm height). The bubble is located at the centre of the cell (axis of symmetry), initially consisting of a truncated sphere of 1 mm radius resting on the electrode with 3/4 of its diameter protruding, which corresponds to an initial contact angle of 60$^\circ$ with the electrode (plane z=0).

Initially the simulation domain is divided into two parts, inside and outside the bubble. The initial values of the phase field are set to -1 and 1 at the interior of the bubble and the surrounding fluid respectively, while the NS equations assume steady flow initial conditions ($\textbf{v}=0$) in the whole domain. The stationary versions of the diffusion equations are solved iteratively using the following initial conditions: the $\text{H}_2$ concentration is set to the value $n_{\text{H}_{2,in}}=n_s=1$ mol/l (inside the bubble) and $n_{\text{H}_{2,out}}=n_{\text{H}_{2r}}=1 $ mol/l (outside the bubble). Also as initial conditions, we assume that there is no proton concentration inside the bubble ($n_{p,in}=0$ mol/l) and that the concentration outside the bubble has a value $n_{p,out}= n_s$ mol/l. Finally, the potential is set to 0 V in the whole domain. At each time step, we solve the stationary versions of the diffusion equations using the values of the previous step. The same approach is followed for the potential.  

For the boundary conditions, non-slip walls ($v_r=0$)  have been used on the contact surfaces of the cell, in which it has been further assumed that the equilibrium contact angle between the gas and liquid phase is $\theta_Y$ through the following relation:
\begin{equation}
	\textbf{n}\cdot \nabla\phi=\cos\theta_Y|\nabla\phi|, 
      \end{equation}
where $\textbf{n}$ refers to the surface normal vector pointing outside the domain.

In the diffusion equations it is further assumed that the side walls of the cell do not diffuse the substances ($\textbf{n}\cdot\nabla n_{\text{H}_2}=0$ ; $\textbf{n}\cdot\nabla n_{\text{n}_p}=0$) and the following Dirichlet boundary conditions are applied: the concentration of $\text{H}_2$ is set to the saturation value on the top surface ($n_{\text{H}_{2,top}}$) of the cell, while the concentration of protons is set to a variable value ($n_{p,top}$). The potential at the top wall is set to $V_{top}=V_0+V$, where $V$ is a controlling variable and $V_0$ is calculated using the following relation:

\begin{equation}
	V_0=\log(n_{p,top}^2/n_{\text{H}_2,top}).
\end{equation}

In addition, the following boundary conditions for the diffusion equations are imposed on the electrode (except the part where the bubble contacts the surface), for the hydrogen equation:

\begin{equation}
	\textbf{n}\cdot \nabla n_{\text{H}_2}= \omega, 
\end{equation}
being $\omega$ the Butler-Volmer constant. The proton equations flux is set to:

\begin{equation}
	\textbf{n}\cdot \nabla n_p= -\omega \delta,
\end{equation}
where $\delta$ is the diffusivity ratio (0.17 in these simulations). Finally the flux imposed in the potential equation is:
\begin{equation}
	n_p\, \textbf{n}\cdot \nabla V= \frac{\partial n_{\text{H}_2}}{\partial z}.
\end{equation}

Finally, the hydrogen production is simulated by injecting a net mass flux into the bubble during the time-dependent analysis, where this flux is calculated in the following way:

\begin{equation}
	J= 1.24\times 10^{-10} \int_{electrode}\frac{\partial  n_{\text{H}_2}}{\partial z} dr.
	\label{eq:massflux}
\end{equation}
The flow inside the bubble diffuses the incoming gas very quickly so that we keep the gas density constant through the whole bubble except for a small neighbourhood around the source.

Lastly, material properties are included for NS physics, setting the dynamic viscosity of water in $\eta_w=0.001 \ \text{Pa}\cdot \text{s}$ and its density in $\rho_w=1000$ kg/m$^3$. The viscosity and density of hydrogen are $\eta_{H_2}=0.00084\, \ \text{Pa}\cdot \text{s}$ and $\rho_{H_2}=0.089$ kg/m$^3$ respectively, where one may notice that hydrogen viscosity is greater than the real value, which was artificially increased in order to guarantee a smooth dispersion of the injected gas and to avoid undesirable side effects, such as excessive phase field distortion produced by vortices in the bubble. Thus the gas rapidly stabilises to a constant density state with zero velocity field. This does not change the overall dynamics of the bubble and important quantities such as the detachment time and gas volume which are the objects of our study. 

The whole domain is meshed using equally-sized triangular elements of linear shape functions with a maximum element length of $0.01$ mm, which gives an average element number of 120000.

\begin{figure}[htbp]
	\centering
	\begin{minipage}{0.45\textwidth}
		\centering
		\includegraphics[width=\linewidth]{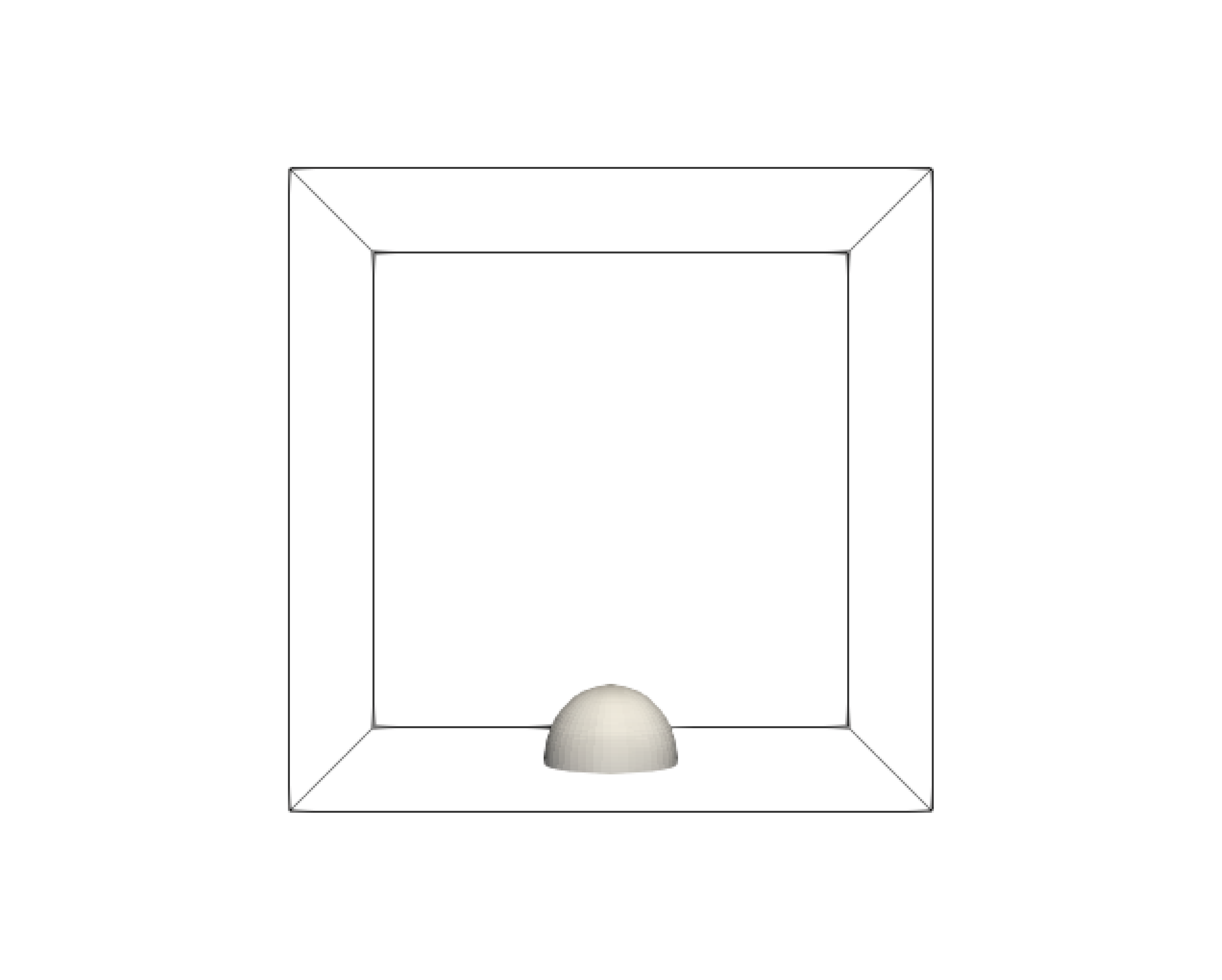}
		\caption*{(a) $t=0.1\,\text{s}$}
	\end{minipage}
	\hfill
	\begin{minipage}{0.45\textwidth}
		\centering
		\includegraphics[width=\linewidth]{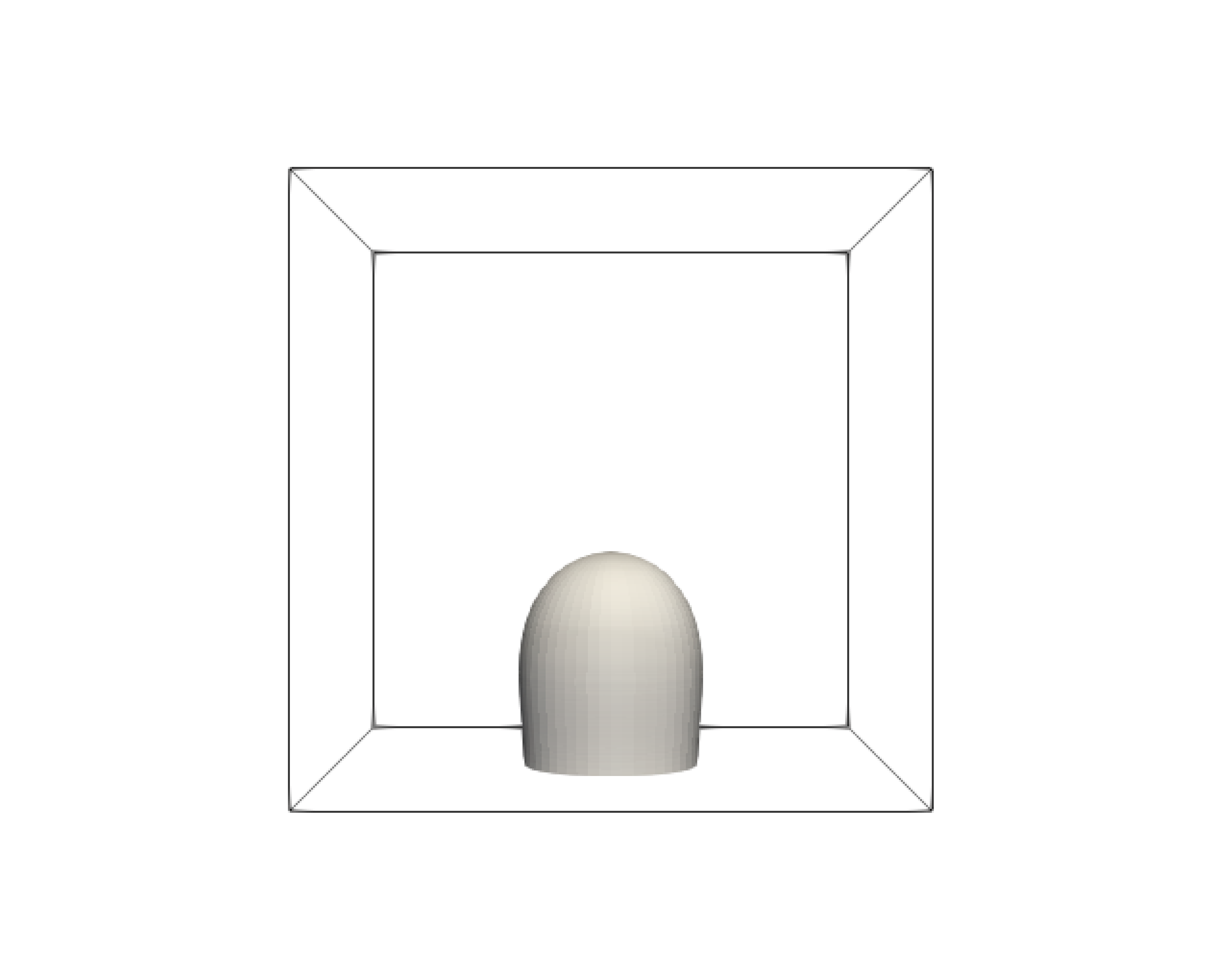}
		\caption*{(b) $t=1.5\,\text{s}$}
	\end{minipage}
	
	\vspace{0.5cm}
	
	\begin{minipage}{0.45\textwidth}
		\centering
		\includegraphics[width=\linewidth]{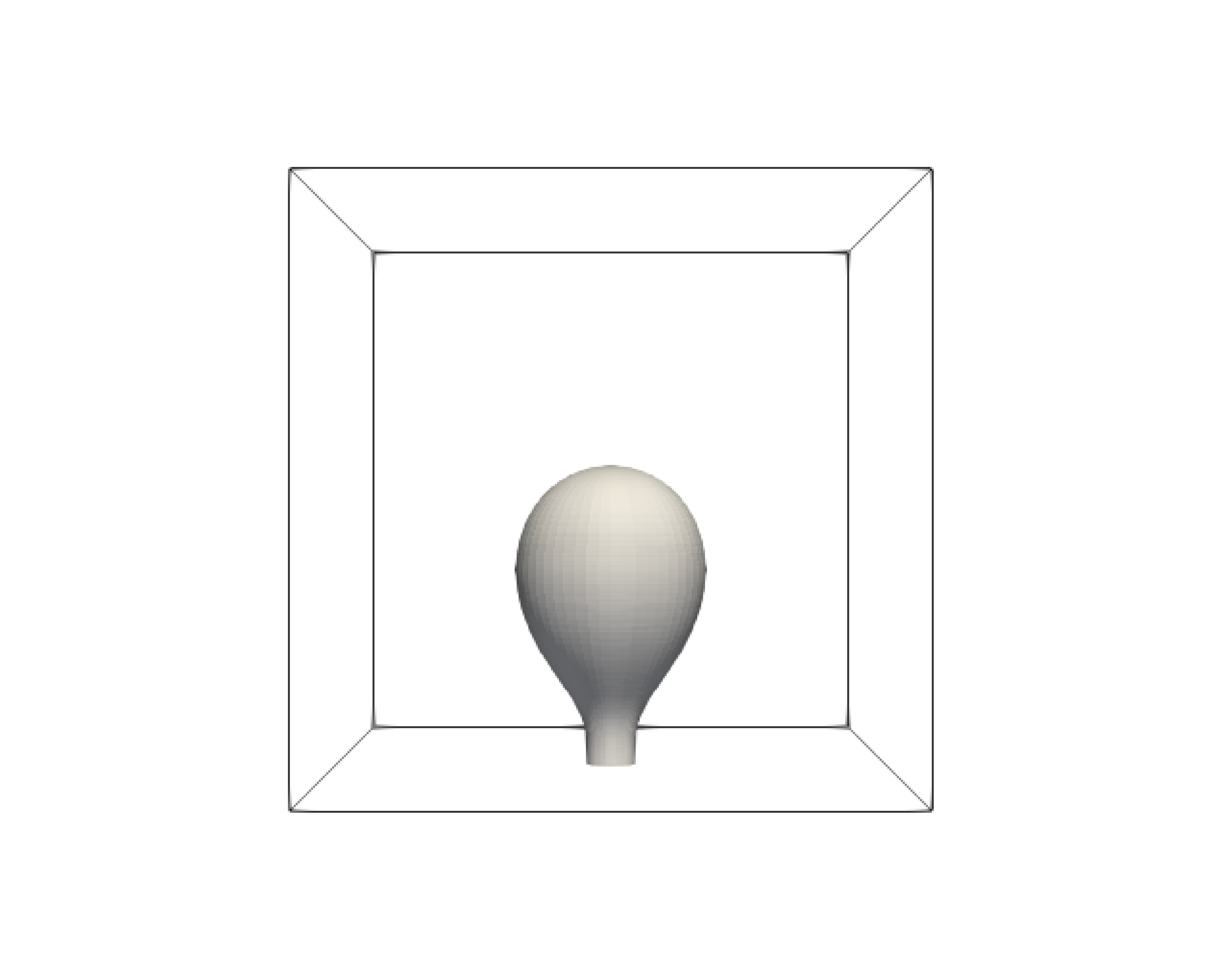}
		\caption*{(c) $t=1.72\,\text{s}$}
	\end{minipage}
	\hfill
	\begin{minipage}{0.45\textwidth}
		\centering
		\includegraphics[width=\linewidth]{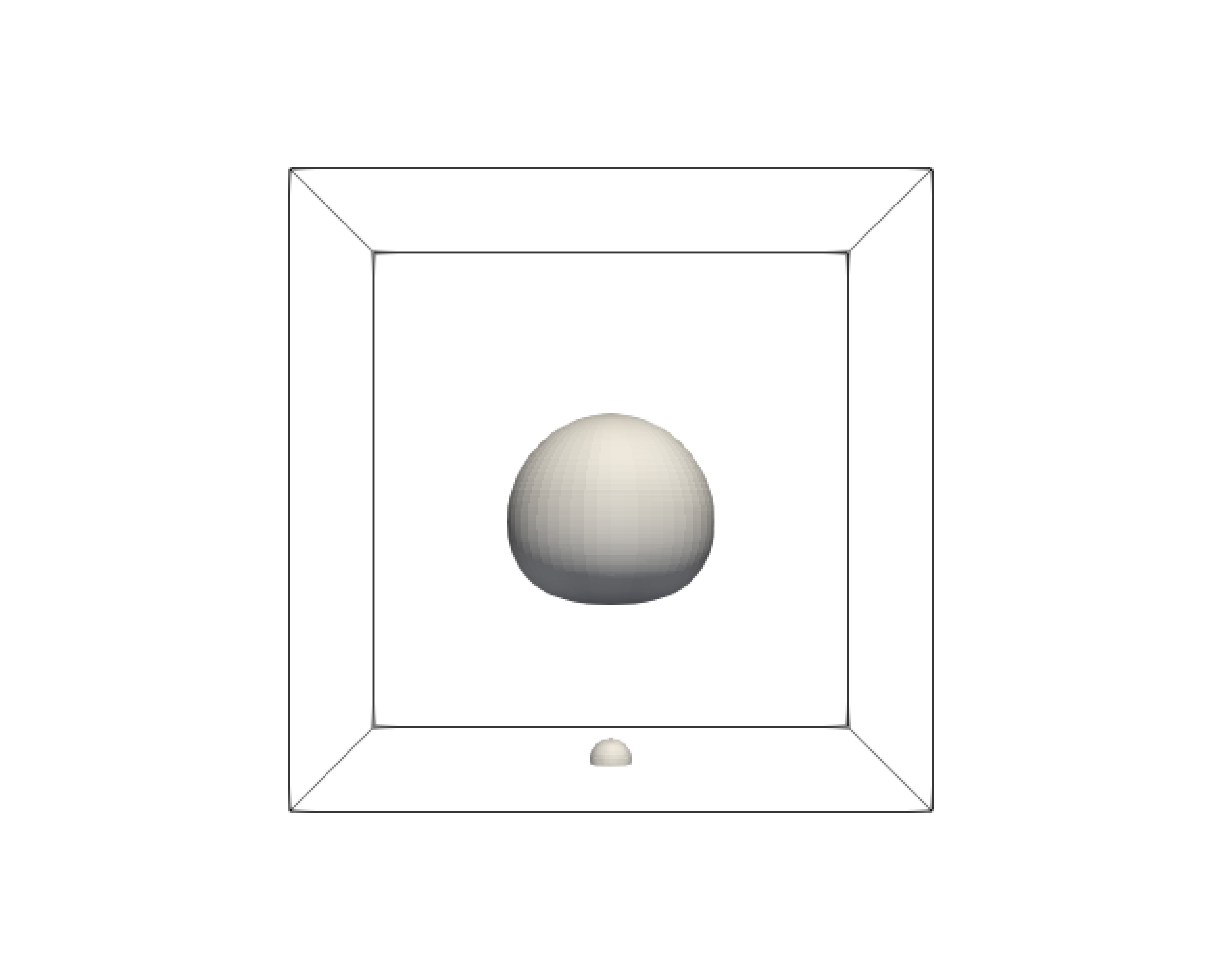}
		\caption*{(d) $t=1.73\,\text{s}$}
	\end{minipage}
	
	\caption{Bubble growth evolution stages for a $\pi/2$ contact angle simulation, an applied voltage $V=5 \, \text{V}$ and the proton concentration at the top surface $n_{p,top}=2\,\text{mol/l}$. The figure shows four stages in the bubble growth. The top-left image shows the initial relaxation to the equilibrium contact angle,while the top right figure shows an intermediate instant where the sphere has grown substantially. The bottom figures shows the instants before and after the lift off, where it may be appreciated the remanent stuck at the electrode produced if the contact angle is great enough.}
	\label{fig:isosurfaces}
\end{figure}

Figure \ref{fig:isosurfaces} shows a simulation with the previously mentioned parameters for a case where the contact angle is $\pi/2$, the voltage applied is $V=5$ V and the proton concentration at the top surface is $n_{p,top}=2$ mol/l.

The figure shows four stages in the evolution of the bubble growth, from left to right and top to bottom.  At the instant $0.1$ s is shown how the angle relaxes from the initial value $\pi/3$ to the equilibrium value $\theta_Y$. The second plot shows the time at $t=1.5$ s, where the bubble has grown considerably. The two last images correspond to the moments right before ($t=1.72$ s) and immediately after ($t=1.73$ s) of the detachment and lift-off. Note that a small amount  of gas has been left on the electrode as a remnant.

\begin{figure}[htbp]
	\centering
	\begin{minipage}{1.0\textwidth}
		\centering
		\includegraphics[width=\linewidth]{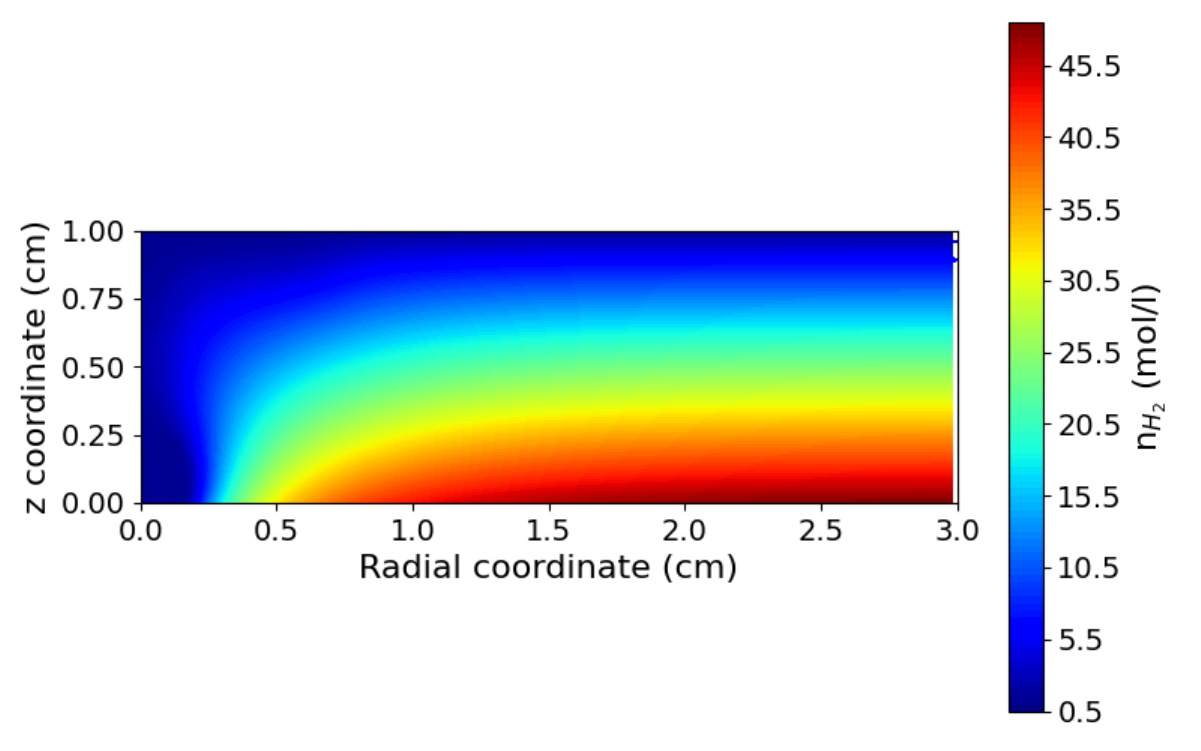}
		\caption*{(a) Hydrogen concentration distribution}
	\end{minipage}
	
	\vspace{0.5cm}
	
	\begin{minipage}{0.9\textwidth}
		\centering
		\includegraphics[width=\linewidth]{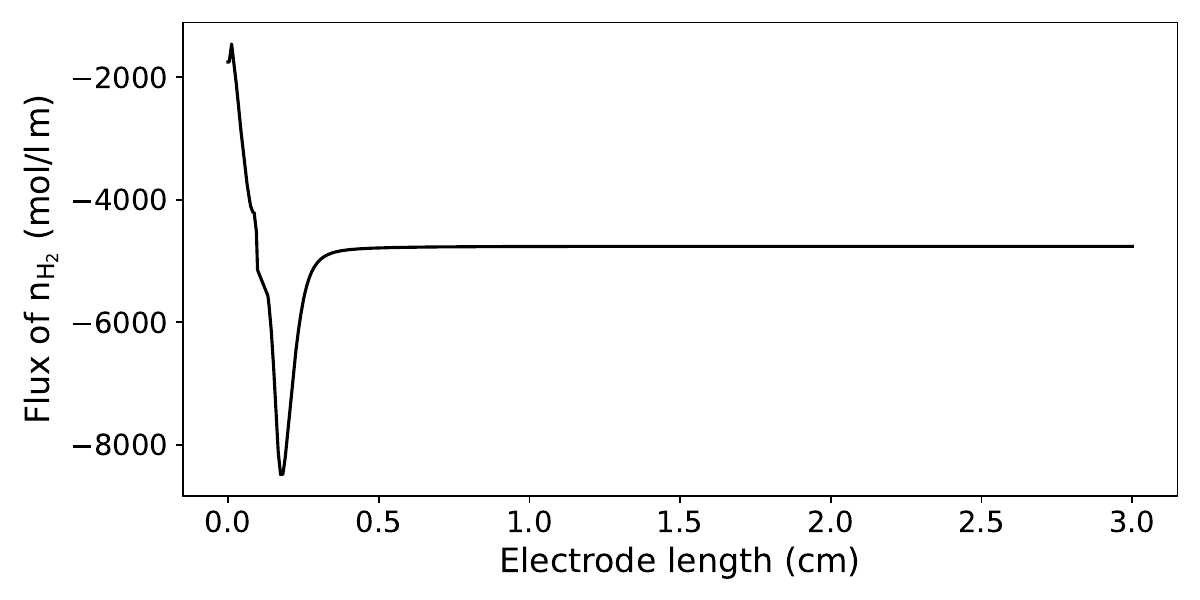}
		\caption*{(b) Hydrogen generation in the electrode}
	\end{minipage}
	
	\caption{Hydrogen generation and distribution at time $t=1$ s prior to bubble lift-off. The top figure (a) presents the distribution map of hydrogen concentration within the cell. The bottom figure (b) displays the spatial derivative of hydrogen concentration with respect to z (vertical flux), calculated along the electrode.}
	\label{fig:concentration}
\end{figure}

An example of hydrogen concentration distribution during the simulations is shown in figure \ref{fig:concentration}. Specifically, the upper part of the figure illustrates the hydrogen concentration for a case with an equilibrium contact angle of $\theta_Y = \pi/2$, at a time step prior to bubble lift-off. It is worth noting that the initial contact angle in the simulations is always  $\pi/3$ and relaxes to the equilibrium angle during the initial stages steps of the simulations. The lower part of the figure depicts the hydrogen flux (i.e. the derivative of the hydrogen concentration $n_{\text{H}_{2}}$ with respect to the vertical coordinate $z$) generated at the electrode and calculated from the corresponding diffusion map. This variation is directly related to the mass flux injected at each time step, as described by equation  (\ref{eq:massflux}).

As expected, the maximum concentration appears in
the neighbourhood of the electrode where the production takes place. The
geometry in the neighbourhood of the contact line is that of a 2D wedge of
angle $\theta _{Y}$. The value of $n_{\text{H}_{2}}$ experiences a jump from its
value at the side of the wedge that corresponds to the electrode to $n_{s}$
at the drop side of the wedge. This implies that a singularity in the
gradient of $n_{\text{H}_{2}}$ must develop as we can see in the lower graph in Figure \ref{fig:concentration}. Hence, an
important part of the production of $\text{H}_{2}$ takes place in the neighbourhood
of the contact line since $\omega =-\partial n_{\text{H}_{2}}/\partial z$
is singular there.

\subsection{The effect of contact angle}

\begin{figure}[htbp]
	\centering
	\includegraphics[width=0.9\textwidth]{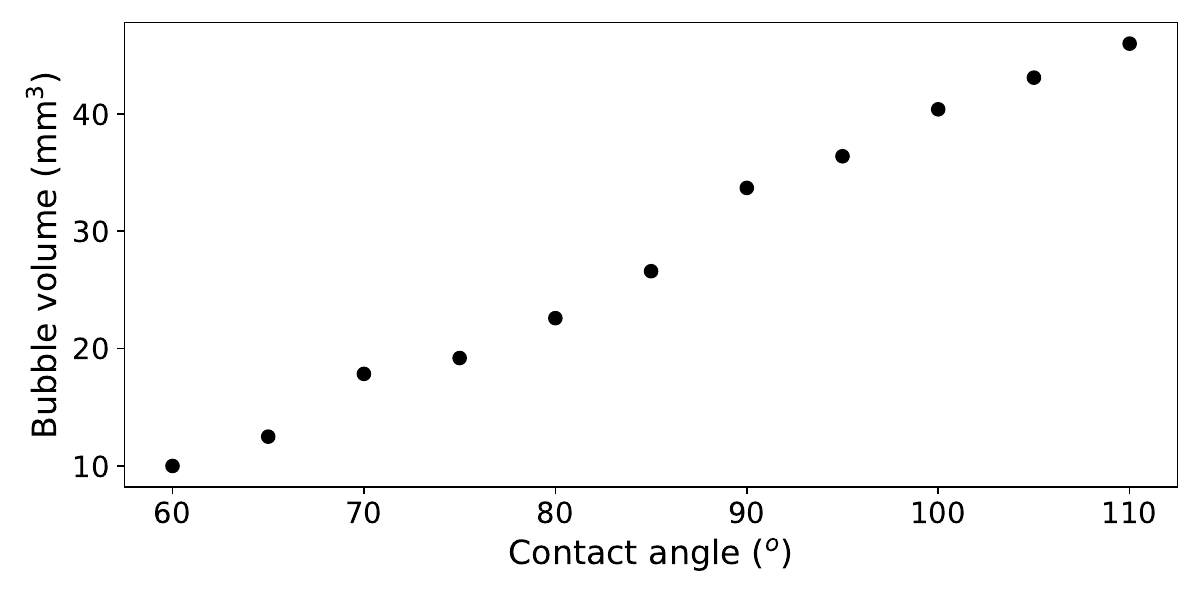} 
	\caption{Bubble volume at lift-off instant for different contact angles. Each point in the graph corresponds to a single simulation were only the contact varies.}
	\label{fig:contact_vol}
\end{figure}

\begin{figure}[htbp]
	\centering
	\includegraphics[width=0.9\textwidth]{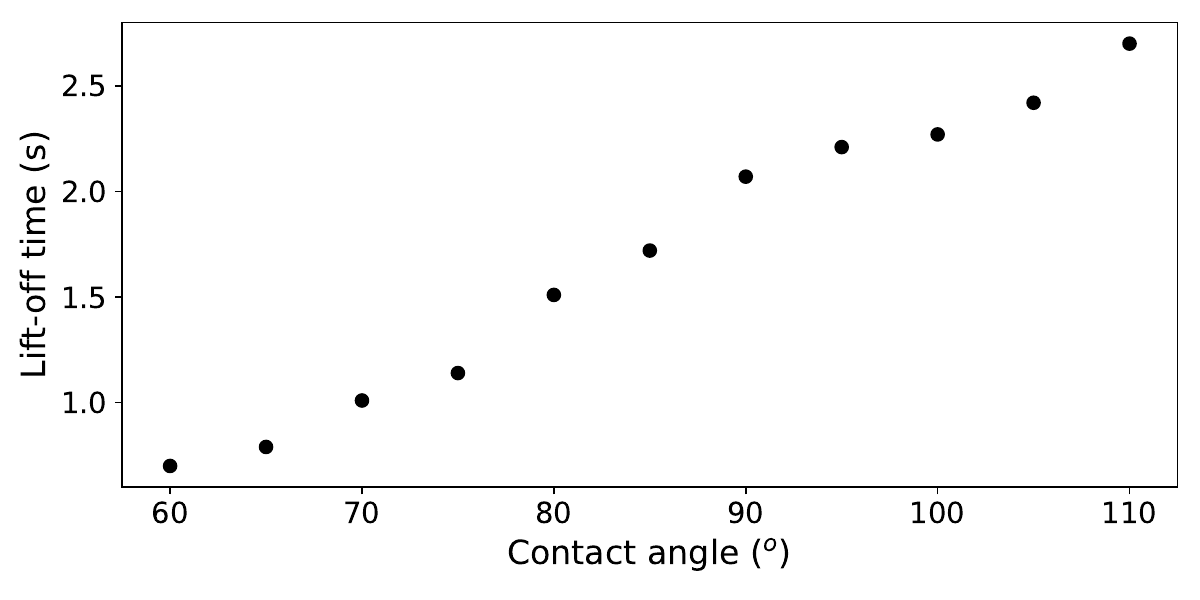} 
	\caption{Time to lift-off for different contact angles. Each point in the graph corresponds to a single simulation were only the contact varies. Overall time is represented in the horizontal axis, which takes into account the coalescence time plus the simulation time from the initial conditions until the moment when the bubbles lifts-off. The coalescence time has been assumed calculating the growth rate in the first instants and extrapolating this trend backwards.}
	\label{fig:contact_time}
\end{figure}

We address next the dependence of the gas production on the contact angle $%
\theta _{Y}$. The size of the bubble when it detaches from the substrate depends strongly on $\theta _{Y}$ as we can see in figure \ref{fig:contact_vol}. For the set of parameters in our simulation, the volume changes a full order of magnitude from $\theta _{Y}=60^\circ$ to $\theta_{Y}=110^\circ$. The smallest bubble, corresponding to $\theta _{Y}=60^\circ$, has a volume of $10$ mm$^{3}$ (i.e. a radius of $1.33$ mm) and the largest, corresponding to $\theta _{Y}=110^\circ$, has a volume of $45.9$ mm$^{3}$ (i.e. a radius of $2.2$ mm). On the other hand, we can evaluate the time to detachment and add the estimated time that would take a nucleating bubble to develop into our initial data in order to estimate the full time that takes for a bubble to grow up to detachment as a function of $\theta _{Y}$. We represent these
times in figure \ref{fig:contact_time}. The time spans from $t=0.7$ s for $\theta _{Y}=60^\circ$ up to $t=2.7$ s for $\theta _{Y}=110^\circ$. Bubbles with $\theta _{Y}=0$ would detach immediately after nucleation with zero volume while bubbles with $\theta _{Y}$ close to $180^\circ$ would take an exceedingly long time to detach. Both of the graphs seem to indicate a linear relationship between the volume of gas produced and the time taken for bubble detachment with respect to the contact angle. Therefore, no optimal angle that maximizes the volume-time ratio can be derived from this analysis. However, it should be emphasized that the volume shown in the graph represents the entire hydrogen gas present in the simulation cell at the moment of lift off, including the remnant that remains attached to the electrode above a certain contact angle ($\theta_Y > 83^\circ$). This remainder can serve as a nucleating core for new bubble growth, speeding up the reaction.

Notice that the trend on volumen of hydrogen at detachment shown in 
Figure \ref{fig:contact_vol} deviates from the parabolic law in our previous article \cite{US} 
and is more like a linear growth. There are several reasons for this. 
On one hand, we are dealing with a dynamical situation as opposed to 
the equilibrium considerations in \cite{US}. Moreover, a balance of energies necessarily has to include, as in electrowetting phenomena, the 
electrostatic energy of the whole system. This is known to produce an 
apparent larger contact angle that is much more noticeable in drops 
with contact angles close to zero.

Our results are consistent with recent experimental findings (see 
\cite{NEW} and references therein). In \cite{NEW}, high-speed imaging and image 
processing were used to analyze bubble growth and detachment in 
various different surfaces. The authors show that superhydrophobic 
surfaces significantly increase bubble volume and reduce formation 
frequency, while superhydrophilic and finely polished surfaces exhibit 
opposite trends.

\subsection{Modifying voltage and acidity}

\begin{figure}[t]
	\centering
	\includegraphics[width=0.9\textwidth]{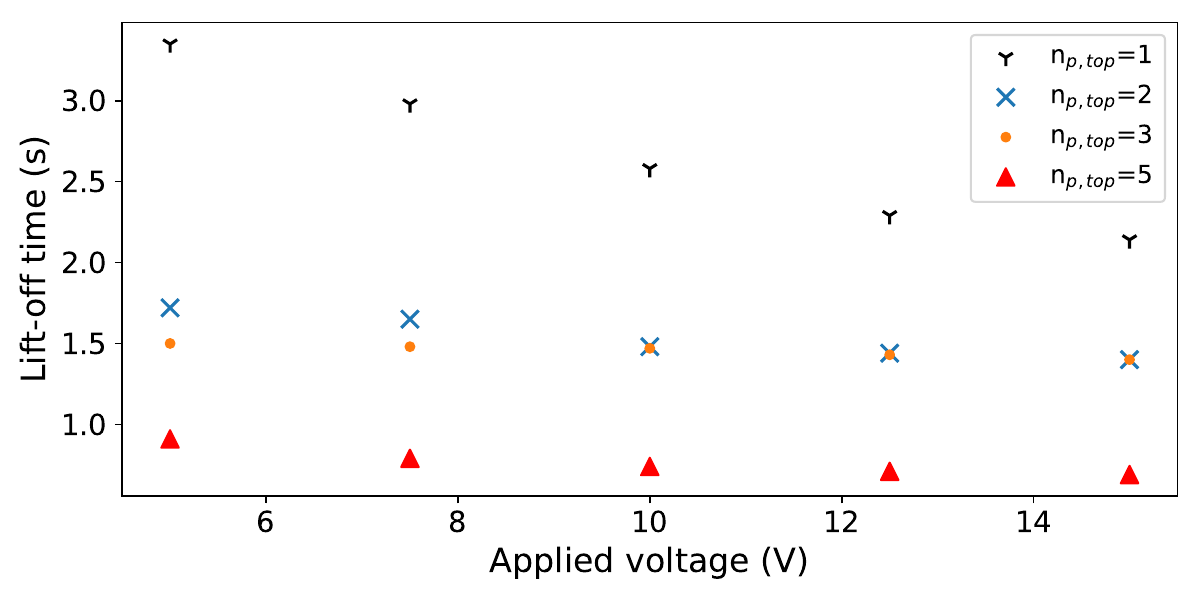} 
	\caption{Lift-off times for different applied voltages. Graphs with different values of reference proton concentrations are shown in the figure, the figure shows the decay of the lift off time respect to the applied voltage. The effect is greater when the proton concentration at the electrode is lower.}
	\label{fig:voltages}
\end{figure}

The two remaining control parameters of the mathematical model presented are the applied voltage at the electrode and the acidity of the medium, which we control by setting the concentration of protons on the top wall of the domain. In this section, we present the most relevant results of this analysis, which can be seen in figure \ref{fig:voltages}, where it is shown how the bubble lift-off time decays with the applied voltage, for different values of the proton concentration. Regarding the latter, it can be seen how the time decay tends to be more pronounced when the acidity of the medium is lower (lower concentration of protons in the upper wall) and remains almost unchanged when the acidity increases. In this figure, the time spans from $t=0.69$ s for $V=15$  V and $n_{p,top}=5$   to $t=3.35$ s  for $V=5$ V and $n_{p,top}=1$. We remark that in all cases a significant increase in the potential does not translate into a equivalent decrease in the detachment time and hence into the hydrogen production. This effect of the applied voltage is even smaller for more acidic media.


\section{Conclusions}
We have studied the growth and detachment of a gaseous bubble in a water electrolyzer. Hydrogen is produced through chemical reactions at an electrode and transported by diffusion into a bubble producing its growth and eventual detachment from the substrate (the electrode). The model consists of fluid mechanical equations (Navier-Stokes) for both the gas inside the bubble and the surrounding fluid, together with classical drift-diffusion equations for the chemical species, i.e. hydrogen molecules and hydrogen ions (protons). The model is completed with an equation for the electric inside the liquid medium. Appropriate boundary conditions have also been introduced, some of them involving the reaction rate at the electrode for the chemical reaction producing H$_2$ from H$^+$. This reaction rate follows the Butler-Volmer equation.

The above model, as such, is very difficult to solve when sharp moving interfaces (between the bubble and the medium) and contact lines are involved. In order to overcome this difficulty we introduce a phase field formulation of the problem and solve it numerically. This has allowed us to perform intensive numerical testing to explore several key aspects: 1) the dependence of the bubble detachment time and volume on the Young's contact angle, 2) the detachment time, for a given Young's angle, on the applied electrical potential, 3) the dependence of the detachment time as a function of the H$^+$ concentration in the medium. We have concluded that the detachment time as well as the amount of gas detached depend strongly on the Young's angle. Nevertheless, if we estimate the gas production rate as the ratio between the detached volume and the detachment time, there is not a substantial dependence on Young's angle. An additional conclusion is that there is not a strong impact of the applied potential on the time of detachment. As a general rule, that time decreases with increasing voltage, but the effect is not strong. In fact, it is almost irrelevant in a medium with very low pH.

In our study we have exploited axial symmetry as a means for reducing the dimensionality of the problem and hence the computation time. The model itself can be easily applied to genuine 3D situations at a higher computational cost. There are various interesting situations to study in such setting: growth of neighbouring bubbles and study of possible screening effects, coalescence of bubble during their growth, etc. We plan to explore these issues in future publications.

\section*{Acknowledgement}
This work has been supported by the TED2021-131530B-I00, PID2022-139524NB-I00 and PID2023-150166NB-I00 projects, as well as URJC 2025/00014/043.

\newpage 


\begin{thebibliography}{99}
  
\bibitem{US} C. Uriarte, M. A Fontelos and M. Array\'as.
Phase field modeling of the detachment of bubbles from a solid substrate. Physics of Fluids 36, 062001 (2024).

\bibitem{Modelo} A. Berm\'udez, P. Font\'an, M. Fontelos, F. Higuera and A. Rivero.
Nucleation, growth and detachment of bubbles formed by reaction
over surfaces. Proceedings of the 163 European Study Group with Industry (163 ESGI) 2021. 

\bibitem{AC} S. M. Allen and J. W. Cahn. A microscopic theory for antiphase
boundary motion and its application to antiphase domain coarsening. Acta
Metallurgica, 27(6):1085-1095, 1979.

\bibitem{CH} J. W. Cahn and J. E. Hilliard, Free energy of a nonuniform
system. I. Interfacial free energy, J. Chem. Phys. 28, 258 (1958).

\bibitem{C} Gunduz Caginalp. An analysis of a phase field model of a free
boundary. Archive for rational mechanics and analysis, 92:205--245, 1986.

\bibitem{DF} Qiang Du and Xiaobing Feng. The phase field method for
geometric moving interfaces and their numerical approximations. Handbook of
Numerical Analysis, 21:425--508, 2020.

\bibitem{J} D. Jacqmin, Calculation of Two-Phase Navier--Stokes Flows Using
Phase-Field Modeling, Journal of Computational Physics, Volume 155, Issue 1,
10 October 1999, Pages 96-127

\bibitem{VCB} V.E. Badalassi, H.D. Ceniceros, S. Banerjee, Computation of
multiphase systems with phase field models, Journal of Computational Physics
190 (2003) 371--397.

\bibitem{EFG} C. Eck, M. Fontelos, G. Gr\"{u}n, F. Klingbeil and O. Vantzos,
On a phase-field model for electrowetting, Interfaces Free Bound. 11 (2009)
259--290.

\bibitem{TQ} T. Qian, X.-P. Wang and P. Sheng, A variational approach to
moving contact line hydrodynamics, J. Fluid Mech. 564 (2006) 333--360.

\bibitem{FGK} M. A. Fontelos, G. Gr\"{u}n, U. Kindel\'{a}n, F. Klingbeil,
Numerical Simulation of Static and Dynamic Electrowetting, Journal of
Adhesion Science and Technology Volume 26, (1805-1824), 2012.

\bibitem{ZLF} Zongliang Zhang, Wei Liu and Michael L. Free, Phase-Field
Modeling and Simulation of Gas Bubble Coalescence and Detachment in a
Gas-Liquid Two-Phase Electrochemical System, Journal of The Electrochemical
Society, 2020 167 013532

\bibitem{JO} R. Jafari and T. Okutucu-\"{O}zyurt, Phase-Field Modeling of
Vapor Bubble Growth in a Microchannel, The Journal of Computational
Multiphase Flows Volume 7, Issue 3, 143-158 (2016)

\bibitem{FA} Farhat, Mohamed ; Avellan, Fran\c{c}ois, On the Detachment of a
leading edge Cavitation, Proceedings of the fourth international Symposium
on Cavitation, Pasadena, Ca, USA, June 2001.

\bibitem{YY} W. Yin, L. Yuan ,H. Huang, Y. Cai, J. Pan, N. Sun, Q. Zhang, Q.
Shu, C. Gu, Z. Zhuang, L. Wang, Strategies to accelerate bubble detachment
for efficient hydrogen evolution, Chinese Chemical Letters 35 (2024) 108351

\bibitem{DGS} H. Ding, M. N. H. Gilani, P. D. M. Spelt, Sliding, pinch-off
and detachment of a droplet on a wall in shear flow, Journal of Fluid
Mechanics, Volume 644 (2010), 217-244

\bibitem{Z} A. Guion , S. Afkhami , S. Zaleski, J. Buongiorno, Simulations
of microlayer formation in nucleate boiling, International Journal of Heat
and Mass Transfer, Volume 127, Part B, December 2018, Pages 1271-1284

\bibitem{FG} J. M. Gordillo and M. A. Fontelos, Satellites in the Inviscid
Breakup of Bubbles, Phys. Rev. Lett. 98, 144503 (2007)

\bibitem{EF} J. G. Eggers, M. A. Fontelos, D. Leppinen, J. H. Snoeijer,
Theory of the collpasing axisymmetric cavity, Physical Review Letters .98,
094502 (2007).

\bibitem{YL} M. Yue, H. Lambert, E. Pahon, R. Roche, S. Jemei, D. Hissel,
Hydrogen energy systems: A critical review of technologies, applications,
trends and challenges, Renewable and Sustainable Energy Reviews, Volume 146,
August 2021, 111180

\bibitem{AFU} C. Uriarte, M. Array\'{a}s, M. A. Fontelos, Phase field
modeling of the detachment of bubbles from a solid substrate, Physics of
Fluids 36, 062001 (2024).

\bibitem{FENG} P. Yue, C. Zhou, J.J Feng. Sharp-interface limit of the Cahn–Hilliard model for moving contact lines. Journal of Fluid Mechanics 645, 279-294 (2010). 
  
\bibitem{COM} COMSOL Multiphysics, v. 6.1., www.comsol.com, COMSOL AB, Stockholm, Sweden.

\bibitem{NEW} H. Rahimi, A. Sattari, M. Mashhadi Keshtiban.Experimental study on 
bubble formation and detachment from submerged orifices with chemically coated and sandpaper-textured surfaces. Colloids and Surfaces A: Physicochemical and Engineering Aspects, Volume 727, Part 2,2025,138317.
\end{thebibliography}
\end{document}